\documentclass{PoS}

\usepackage{braket}
\usepackage{subfig}
\usepackage{float}
\usepackage{url}
\usepackage{amsmath}
\usepackage{amsfonts}
\usepackage{verbatim}
\usepackage{listings}
\usepackage{algorithm}
\usepackage{algorithmic}

\newcommand{\Tr}{\mbox{\rm Tr}}
\newcommand{\Real}{\mbox{\rm Re}}

\title{Landau gauge fixing on the lattice using GPU's}

\ShortTitle{Landau gauge fixing on the lattice using GPU's}

\author{Nuno Cardoso \\
CFTP, Departamento de F\'{\i}sica, Instituto Superior T\'ecnico,
Universidade T\'ecnica de Lisboa, Avenida Rovisco Pais 1, 1049-001 Lisbon, Portugal\\
E-mail: \email{nuno.cardoso@ist.utl.pt}}
\author{\speaker{Paulo J. Silva} \\
CFC, Departamento de F\'{\i}sica, Faculdade de Ci\^encias e Tecnologia, Universidade de Coimbra, 3004-516 Coimbra, Portugal\\
E-mail: \email{psilva@teor.fis.uc.pt}}
\author{Orlando Oliveira\\
CFC, Departamento de F\'{\i}sica, Faculdade de Ci\^encias e Tecnologia, Universidade de Coimbra, 3004-516 Coimbra, Portugal\\
E-mail: \email{orlando@teor.fis.uc.pt}}
\author{Pedro Bicudo\\
CFTP, Departamento de F\'{\i}sica, Instituto Superior T\'ecnico,
Universidade T\'ecnica de Lisboa, Avenida Rovisco Pais 1, 1049-001 Lisbon, Portugal\\
E-mail: \email{bicudo@ist.utl.pt}}

\abstract{In this work, we consider the GPU implementation of the steepest descent method with Fourier acceleration for Laudau gauge fixing, using CUDA. The performance of the code in a Tesla C2070 GPU is compared with a parallel CPU implementation.}

\FullConference{Xth Quark Confinement and the Hadron Spectrum\\
		 8-12 October 2012\\
		 TUM Campus Garching, Munich, Germany}

\begin{document}

On the lattice, Landau gauge is defined through the maximization of the functional
	$$	F_U[g]=\frac{1}{N_d N_cV}\sum_x\sum_\mu\Real\left[ \Tr\left(  g(x)U_\mu(x)g^\dagger(x+\mu) \right) \right] \, ,$$
where $N_d$ is the dimension of the space-time, $N_c$ is the dimension of the gauge group and $V$ the lattice volume, on each gauge orbit. The functional $F_U[g]$ can be maximised using a steepest descent method \cite{Davies:1987vs,Oliveira:2003wa}. However, when the method is applied to large lattice volumes, it  faces the problem of critical slowing down, which can be attenuated by Fourier acceleration.

The main goal of this work is to compare the difference in performance between GPU and CPU implementations of the Fourier accelerated Landau gauge fixing method. The MPI parallel version of the algorithm was implemented in C++, using the machinery provided by the Chroma library \cite{Edwards2005}; for the Fourier transforms, the code uses PFFT, a parallel FFT library written by Michael Pippig \cite{Pippig2011}. For the GPU code, we used version 4.1 of CUDA  \cite{NVIDIA:cudac} -- see also \cite{Cardoso:2011xu}; FFT are performed using the CUFFT library by NVIDIA \cite{CUFFT}.

For such a comparison, we use a NVIDIA Tesla C2070. The GPU code has been run using a 12 real number representation; furthermore, we used texture memory and we switched ECC off. The CPU code has been run in the Centaurus cluster, at Coimbra. In Centaurus, each node has 2 Intel Xeon E5620@2.4 GHz (quad core), with 24 GB of RAM, and it is equipped with a DDR Infiniband network. 

In order to compare the performance of the two codes, we used a $32^4$ lattice volume. The configurations have been generated using the Wilson gauge action, with three different values of $\beta$. The runs used $\alpha=0.08$ and $\theta < 10^{-15}$. 

\begin{figure}[!htb]
\begin{center}
\includegraphics[width=10cm]{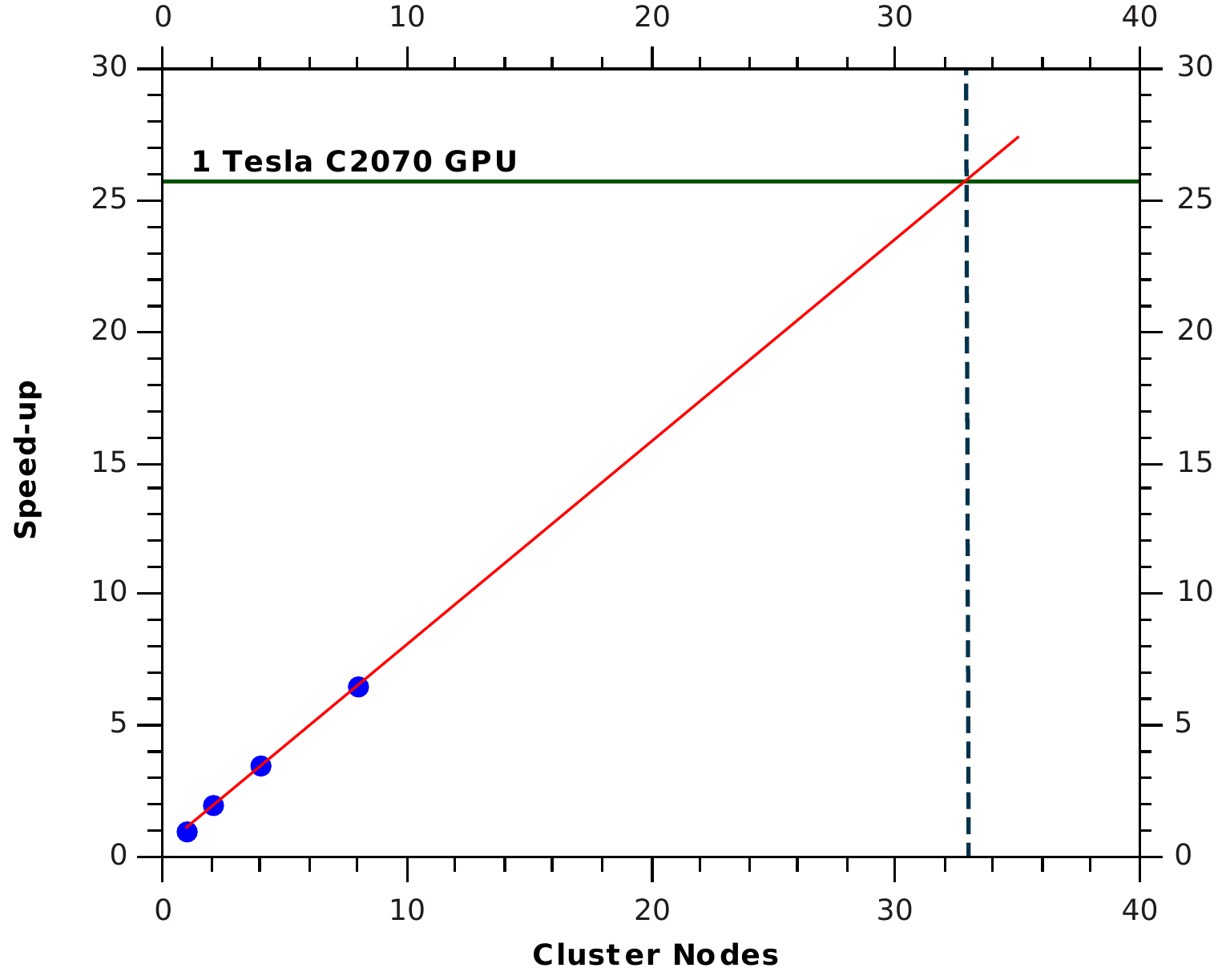}
\end{center}
\caption{Strong scaling CPU tested for a $32^4$ lattice volume and comparison with the GPU for the best performance, 12 real number parametrization, ECC Off and using texture memory in double precision, \cite{Cardoso:2012pv}. In Centaurus, a cluster node means 8 computing cores.}
\label{ss32}
\end{figure}

In figure \ref{ss32} we compare the performance of the parallel CPU code against the GPU result. The CPU code shows a good strong scaling behaviour, with a linear speed-up against the number of computing nodes. However, the GPU code was much faster: in order to reproduce the performance of the GPU code, one needs 256 CPU cores.

For more details on this work, please see 
\cite{Cardoso:2012pv, Cardoso:2012ur, Cardoso:2012uu}.

\section*{Acknowledgments}

We thank B\'{a}lint Jo\'{o} and Michael Pippig for discussions about Chroma and PFFT libraries respectively. In particular, we thank Michael Pippig for extending his library for 4-dimensional FFTs. 

This work was partly funded by the FCT contracts  POCI/FP/81933/2007, 
CERN/FP/83582/2008, PTDC/FIS/100968/2008, CERN/FP/109327/2009, CERN/FP/116383/2010, CERN/FP/123612/2011, projects developed under initiative QREN financed
by UE/FEDER through Programme COMPETE.
Nuno Cardoso and Paulo Silva are supported by FCT under contracts SFRH/BD/44416/2008 and SFRH/BPD/40998/2007 respectively.
We would like to thank NVIDIA Corporation for the hardware donation used in this work via Academic Partnership 
program.

\bibliographystyle{elsarticle-num}
\bibliography{bib}

\end{document}